\documentclass{article}
\usepackage{spconf,amsmath,graphicx}

\usepackage{algorithm}
\usepackage{algorithmic}

\usepackage{booktabs}  
\usepackage{refcount}
\usepackage{bbm}
\usepackage{makecell}
\usepackage{amssymb}
\usepackage{amsthm}
\usepackage{xcolor}

\usepackage{multirow}
\usepackage{xspace}
\usepackage{subcaption}
\usepackage{hyperref}
\usepackage{newfloat}
\usepackage{listings}
\usepackage{pifont}
\newcommand{\cmark}{\ding{51}}%
\newcommand{\xmark}{\ding{55}}%

\newcommand{\PreserveBackslash}[1]{\let\temp=\\#1\let\\=\temp}
\newcolumntype{C}[1]{>{\PreserveBackslash\centering}p{#1}}

\definecolor{dred}{rgb}{0.8, 0.0, 0.2}

\hypersetup{
  colorlinks,
  citecolor=black,
  linkcolor=black,
  urlcolor=blue,
  pdftitle={paper}
}

\title{Designing and Evaluating Speech Emotion Recognition Systems: A reality check case study with IEMOCAP}
\name{Nikolaos Antoniou $^{\dagger,\ast}$, Athanasios Katsamanis $^{\dagger,\ast}$, Theodoros Giannakopoulos $^{\dagger}$, Shrikanth Narayanan $^{\dagger,+}$}

\address{
$^\dagger$ Behavioral Signal Technologies, Los Angeles, CA, USA\\
$\ast$ Institute for Language and Speech Processing, Athena Research Center, Athens, Greece \\
$+$ SAIL-University of Southern California, Los Angeles, CA, USA \\
\small \texttt{\{nikos.antoniou,nassos,thodoris,shri\}@behavioralsignals.com}}
\begin{document}
%
\maketitle

\begin{abstract}

There is an imminent need for guidelines and standard test sets to allow direct and fair comparisons of speech emotion recognition (SER).  
While resources, such as the  \emph{Interactive Emotional Dyadic Motion Capture} (IEMOCAP) database, have emerged as widely-adopted reference corpora for researchers to develop and test models for SER, published work reveals a wide range of assumptions and variety in its use that challenge reproducibility and generalization. Based on a critical review of the latest advances in SER using IEMOCAP as the use case, our work aims at two contributions: First, 
 using an analysis of the recent literature, including assumptions made and metrics used therein, we provide a set of SER evaluation guidelines. Second, using recent publications with open-sourced implementations, we focus on reproducibility assessment in SER. 


\end{abstract}


\section{Introduction} \label{sec:Intro}

Systems with the ability to recognize emotions, simply by processing speech information can be particularly useful for building intelligent machines that can incorporate perceived affective expressions. Inferring expressed human affective state  can, for instance, help a voice assistant to adjust its response to the user. Speech Emotion Recognition (SER) falls within a broad class of problems within \emph{computational paralinguistics} \cite{Schuller2013ParalinguisticsinSpeechand}. 


As is typical in such speech machine-learning applications, the cornerstone of progress is the use of annotated datasets. These resources enable us to build and evaluate the performance of systems that learn how to map successfully an input speech sample to the desired output. Collecting appropriate input-label pairs for the SER task is a highly ambiguous procedure. The ambiguity lies on the fact that the quality of labels depends on the annotators' nuanced perception of emotions. 
Besides that, there is an inherent trade-off between data quality and the desire to facilitate annotation when gathering emotional speech. Spontaneously spoken segments, for example, can be of high (audio) quality but may contain overlapping and or ambiguous affective expressions which can confuse annotators and exacerbates the difficulty of labelling such data \cite{Mower2009Interpretingambiguousemotionalexpressions}. On the other hand, imposing a one-to-one correspondence between a certain speech segment and its emotion may give rise to overly artificial data samples for learning and inference.  

Several published studies in the SER domain have used 
the \emph{Interactive Emotional Dyadic Motion Capture} (IEMOCAP) \cite{iemocap} database. Using a critical review of the recent works using IEMOCAP for SER as use cases, this work aims to investigate questions centered on how the adopted evaluation approaches compare across the studies, and how guidelines for comparable evaluation and reproducible findings can be developed. 
Our contributions are summarized as follows: 
\begin{itemize}
    \item Using a critical review of past studies, we identify three methodological limitations in the adopted evaluation methods. 
    Our analysis reveals that the lack of consensus about the evaluation protocol leads to results of narrowed range which can not be properly contextualized and generalized across related work in the domain. Based on these insights, we summarize a recommended set of evaluation guidelines (building on practice from several previous works), with the end goal of minimizing potential blind spots related to the overall SER procedure.
    \item We also report a reproducibility study, assessing whether models which provide opensource code implementations deliver results consistent with the published findings.  
\end{itemize}
\section{Related Work} \label{sec:RelatedWork}
Deep Learning approaches have triggered a paradigm shift including speech emotion recognition (SER) research and development.  
Systems that solve SER have largely transitioned from classical machine learning models, such as HMMs or SVMs, applied on top of hand engineered features to end-to-end learning of Deep Neural Networks (DNNs). 

In designing deep-learning SER systems, the practitioner faces a series of important choices. The first relates to the way that speech information is represented, e.g., using MFCCs, spectrograms or other acoustic features. Then, it is crucial to identify a suitable architecture that will encode this information efficiently, with the most natural candidates being CNNs \cite{aftab22lightsrnet,zhu2022speech,xu2021speech}, LSTMs \cite{santoso21_interspeech,feng20_interspeech}, or even a combination \cite{li19n_interspeech,zou2022speech} of these. 

Relatedly, Self-Supervised Learning (SSL) features have received a mounting interest in the last years.
There is a flurry of works exploring the prospect of using wav2vec \cite{schneider2019wav2vec,baevski2020wav2vec} or HuBERT \cite{hsu2021hubert} features in emotion recognition \cite{wang2021fine,gat2022speaker,pepino21_interspeech,yang21c_interspeech,cai21b_interspeech}. On a similar vein, Conformer Applied to Paralinguistics (CAP) \cite{shor2022universal} are representations based on the Conformer \cite{gulati2020conformer} architecture, trained similarly to the wav2vec-2.0 (w2v2) model, with their downstream performance being evaluated on the Non-Semantic Speech Benchmark (NOSS) \cite{shor20_interspeech}. 
In \cite{shor2022trillsson}, authors have distilled the efficient CAP representations to more lightweight architectures, aiming to significantly reduce memory and compute overheads, while sacrificing the downstream performance only slightly. 

On a different note, it is critical to associate our work with prior publications, such that of Musgrave et al. \cite{musgrave2020reality}, that aim to increase awareness about  methodological flaws found in ML papers. Such works contribute in improving the experimental rigor of the field, and play a key role in realizing the potential limitations related to the reported results of each method.   
\section{Revisiting Iemocap} \label{sec:IEMOCAP}

\subsection{Dataset Details}

The IEMOCAP \cite{iemocap} database consists of five dyadic interactions sessions, each between a unique male and female speaker per session,  amounting to a total of 10 speakers. The conversation of each session is segmented based on speaker turns, and these conversational segments are annotated for perceived expressed emotions. Each segment is labelled by 3 different annotators, where they assigned both a discrete categorical label (e.g. happy, neutral, sad etc.) and a continuous valued one, assessing the valence, dominance and activation dimensions. Additionally, the conversations are a blend of both scripted and improvised speech interactions.  

Although the diversity of label information present in this database enables the study of SER from multiple aspects, our focus in this paper is on the most prevalent setting found across literature. In this setting, SER is tackled as a 4-way classification problem, discarding all conversational segments whose discrete labels are not included in the following set \{\emph{neutral}, \emph{happy}, \emph{sad}, \emph{angry}, \emph{excited}\}. Then, due to their expressive closeness, the happy and excited label classes are merged. 
This process leads to a dataset of $N=5531$ samples, with a distribution of \{1708, 1636, 1103, 1084\} examples for the \{\emph{neutral}, \emph{happy}, \emph{sad}, \emph{angry}\} classes, respectively. After the filtering procedure, the dataset ends up with audio segments of average duration of $\sim 4.5$ seconds, leading to 7 hours of speech data, coming from 10 different speakers.

The resulting dataset comes with two major challenges: First, it contains a limited amount of speech, compared to other datasets used in SER e.g., \emph{MSP-Podcast} \cite{lotfian2017building} amounts to 27 hours. Second, the empirical label distribution is imbalanced, with most samples originating from the neutral and happy classes. Both of these challenges affect machine learning methods in various ways; the former means that models of increased capacity suffer from overfitting which should be treated by strong regularization, e.g., with multitask learning \cite{li19n_interspeech,cai21b_interspeech,gat2022speaker}. Failing to properly address the latter issue will probably introduce unsolicited biases to the system's outputs. Recent works propose elaborate data augmentation techniques \cite{Chatziagapi2019,pappagari2021copypaste} or the use of different losses, e.g., the focal loss in the case of Aftab et al. \cite{aftab22lightsrnet}, to combat imbalance. 

\subsection{Evaluation Protocol}
In this section, we highlight the three main assumptions typically made by researchers during the evaluation of their SER models. This leads to a vast amount of distinct evaluation protocols, with final results that cannot be compared easily or fairly with previous methods in the literature. After that, we present a set of a evaluation guidelines, hoping to disambiguate the evaluation process and mitigate the errors made therein. 

\begin{table*}[!ht]
    \footnotesize
    \centering
    \begin{tabular}{cccccccc}
    \toprule
       & & & & & & & \\ 
       \# & Publication & Code & Acoustic Feats & UA (\%) & WA (\%) & cross-val & Comments on Eval.\\
      \cmidrule(lr){1-1} \cmidrule(lr){2-2} \cmidrule(lr){3-3}  \cmidrule(lr){4-4} \cmidrule(lr){5-6} \cmidrule(lr){7-7} \cmidrule(lr){8-8} 
      1 & Feng et al.~\cite{feng20_interspeech} & \href{https://github.com/Kyoto-University-Speech-and-Audio/feng-asr-ser}{Yes} & MFCCs & 76.4 & 75.5 & 5-fold & -\\
      2 & Zhu and Li~\cite{zhu2022speech}       & \href{https://github.com/lixiangucas01/GLAM}{Yes} & MFCCs & 73.90 & 73.70 & 5-fold  & Excited as Happy\\
      3 & Xu et al.~\cite{xu2020improve}        & \href{https://github.com/lessonxmk/head_fusion}{Yes} & Spectrograms  & 67.94 & 67.28 & 5-fold  & Excited as Happy \\
      4 & Gat et al.~\cite{gat2022speaker}      & No & HuBERT & - & 81.0 & 5-fold & -\\
      5 & \multirow{2}{*}{Wang et al.~\cite{wang2021fine}}  & \multirow{2}{*}{\href{https://github.com/speechbrain/speechbrain/tree/develop/recipes}{Yes}}        & HuBERT & -  & 79.58  & \multirow{2}{*}{5-fold} & - \\
      6 &                                                   &                             & w2v2   & -  & 77.47 &  & -\\ 
      7 & Peng et al.~\cite{peng2021efficient} & No & MFCCs & 79.1 & 78.0 & 10-fold & -\\ 
    \cmidrule{1-8}
      8 & Xu et al.~\cite{xu2021speech}        & \href{https://github.com/lessonxmk/Optimized_attention_for_SER}{Yes} & Spectrograms  & 77.54 & 79.34 & 5-fold  & Improv. only, Exc. as Hap.\\
      9 & Xu et al.~\cite{xu2020improve}       & \href{https://github.com/lessonxmk/head_fusion}{Yes}                 & Spectrograms  & 76.36 (63.92) & 76.18 (65.90) & 5-fold  & Improv. (Scripted) only, Exc. as Hap. \\
      10 & Zhu and Li~\cite{zhu2022speech}     & \href{https://github.com/lixiangucas01/GLAM}{Yes}                    & MFCCs & 79.25 (70.39) & 81.18 (71.44) & 5-fold  & Improv. (Scripted) only, Exc. as Hap. \\
      11 & Liu and Wang~\cite{liu21n_interspeech}  & No                                                                   & MFCCs & 78.30 & 79.52 & 5-fold  & Improv. only \\
      12 & Moine et al.~\cite{moine21_interspeech} & No                                                                   & Spectrograms  & 77.22 & -     & 5-fold  & Improv. only\\
    \bottomrule
    \end{tabular}
    
    \caption{Comparison of methods that perform Speaker Dependent (SD) evaluation. Rows 8-12 perform random 5-fold cross val. only on the improvised data. }
    \label{table1}
\end{table*}
\vspace{-0.1cm}
\begin{table*}[!ht]
    \footnotesize
    \centering
    \begin{tabular}{ccccccc}
    \toprule
      & & & & & & \\ 
      \# & Publication & Code & Acoustic Feats & UA (\%) & WA (\%) & cross-val\\
      \cmidrule(lr){1-1} \cmidrule(lr){2-2} \cmidrule(lr){3-3}  \cmidrule(lr){4-4} \cmidrule(lr){5-6} \cmidrule(lr){7-7}  
      1 & Pepino et al.~\cite{pepino21_interspeech} & \href{https://github.com/habla-liaa/ser-with-w2v2}{Yes} & w2v2 & 67.2 & - & 5-fold \\
      2 & Yang et al.~\cite{yang21c_interspeech}    & \href{https://github.com/s3prl/s3prl}{Yes} & HuBERT & 67.62 & -    &  5-fold \\
      3 & Gat et al.~\cite{gat2022speaker}          & No & HuBERT & -     & 74.2 & 5-fold \\
      4 & Santoso et al.~\cite{santoso21_interspeech}  & No & MFCC+CQT+F0 & 75.9 & 76.1   & 5-fold \\
      5 & Li et al.~\cite{li19n_interspeech}       & No & Spectrograms & 82.8 & 81.6 & 5-fold \\
      6 & \multirow{2}{*}{Zou et al.~\cite{zou2022speech}} &  \multirow{2}{*}{\href{https://github.com/vincent-zhq/ca-mser}{Yes}} &  \multirow{2}{*}{MFCCs,Spec,w2v2} & 71.05 & 69.80 & 5-fold \\ 
      7 & & & & 72.70 & 71.64 & 10-fold \\ 
      8 & \multirow{2}{*}{Wang et al.~\cite{wang2021fine}} & \multirow{2}{*}{\href{https://github.com/speechbrain/speechbrain/tree/develop/recipes}{Yes}} & HuBERT & - & 73.01 & \multirow{2}{*}{10-fold}  \\
      9 &                                                  &                                                                                              & w2v2   & - & 70.99 &  \\ 
      10 & Aftab et al.~\cite{aftab22lightsrnet} & \href{https://github.com/AryaAftab/LIGHT-SERNET}{Yes} & MFCCs & 70.76 & 70.23 & 10-fold \\
      11 & Feng et al.~\cite{feng20_interspeech} & \href{https://github.com/Kyoto-University-Speech-and-Audio/feng-asr-ser}{Yes} & MFCCs & 69.67 & 68.63 & 10-fold \\
      12 & Cai et al.~\cite{cai21b_interspeech}  & \href{https://github.com/TideDancer/interspeech21_emotion}{Yes} & w2v2 & - & 78.15 & 10-fold \\
      13 & Shor et al.~\cite{shor2022universal}  & No                                                  & CAP & - & 79.2  & Session 05  \\
      14 & Shor and Venugopalan~\cite{shor2022trillsson} & \href{https://tfhub.dev/s?q=trillsson}{Yes} \hyperref[trill]{\dag}  & \textsc{TRILLsson} & - &  73.2 & Session 05 \\
    \bottomrule
    \end{tabular}
    
    \caption{Comparison of methods that perform Speaker Independent (SI) evaluation. The papers of rows 13,14 use only Session05 (i.e. speakers '05M' and '05F') as test set. \label{trill}(\dag): They only provide the trained weights of the feature extractors. }
    \label{table2}
\end{table*}

\subsubsection{Speaker Dependent Evaluations}
In speech related applications with multiple speakers, it is crucial to guarantee that the speakers on test data are not present in the training set. Otherwise, the system may deceptively appear to be highly performing, whereas in reality it may be the case that it only learns to exploit spurious correlations in an optimal way.  For this reason, the evaluation procedure should adhere to the property of Speaker Independence (SI). 

An important challenge associated with IEMOCAP is that it lacks an established nominal test set; hence, in each work researchers are left with some freedom on how to carry out evaluation. There are two prevalent ways to perform SI evaluations on IEMOCAP: (1) 5-fold cross validation (or leave-one-session-out), where in each turn four sessions (with eight speakers) are used as training data and one session (with two speakers) as test data and (2) 10-fold cross validation (or leave-one-speaker-out), where one speaker is kept for test and the other nine for training set. Usually, two speakers are preserved for validation set purposes in 5-fold cross-validation (one speaker for the 10-fold case). The per-fold accuracy metrics are aggregated through the Weighted Accuracy (WA) and Unweighted Accuracy (UA) metrics. The WA measures the percentage of correct predictions, whereas UA averages the recall metric for each class.   

Intriguingly, the necessity of performing SI evaluation is yet to become clear in the community, since many recent works randomly create the train-val-test split with overlapping speaker identities. This, in contrast to SI, is called Speaker Dependent (SD) evaluation. Besides the obvious violation of speaker independence, the random split limits the possibility of comparisons only to the very specific baselines used in the respective works, and limits generalization.

\subsubsection{Evaluation on Improvised Interaction Data}

In our previous analysis, we mentioned that IEMOCAP contains speech from both improvised and scripted interactions. Another common evaluation trend is to assess performance on only one of these two components. This frequently happens with the improvised component, e.g.,  \cite{moine21_interspeech,xu2021speech}.
While this choice may be linked with the unique goals of each paper, it should be noted that this practice may again impose limits on our ability to contextualize results with other published methods. A desirable solution to this challenge would be to perform -- and make available -- evaluation on both improvised and scripted data, and report results about each data part separately.


\subsubsection{The `excitement' class case}

Here, we discuss another widely-adopted choice which leads to the existence of a new evaluation branch that limits broad comparisons. There are some early works \cite{chernykh2017emotion,tarantino19_interspeech} which only consider the \{\emph{neutral}, \emph{sad}, \emph{angry}, \emph{excited}\} classes, excluding the \emph{happy} class from their study (instead of merging it with the \emph{excited} class). This, in turn, had prompted some subsequent papers \cite{xu2020improve,xu2021speech,zhu2022speech} to conduct evaluation following this protocol. 
Similarly with the two aforementioned evaluation challenges, we believe that researchers should follow the convention of including the merged \textit{happy} and \textit{excited} classes, and then include comparisons with specific baselines that, for example, use only the \emph{excited} class. 

\subsubsection{Comparing literature results}

We collect the reported results of multiple recent SER methods developed on the IEMOCAP database. In Tables \ref{table1} and \ref{table2}, we demonstrate the reported performance metrics for works that use SD and SI evaluations, respectively. We also include additional information, e.g., how the speech signal is represented in each work, whether the respective paper has an official implementation or any other choices followed during evaluation. In the SD case, the results cannot be compared fairly since data are split randomly (except the case that some work evaluates other methods in the exact same partition approach, e.g., \cite{zhu2022speech} outperforms \cite{xu2020improve,xu2021speech}). Overall, it becomes evident that diverging from (and absent) a common evaluation setting can result in an unclear situation and confusion about where each method stands within the published literature, and can impede progress as a community.

\subsection{Recommended evaluation guidelines}\label{eval}

Our proposal for future work on SER using IEMOCAP 
is to perform evaluations according to the following protocol to ensure a common minimally-viable and comparable baseline:
\begin{itemize}
    \item Use \emph{neutral, sad, angry} and \emph{happy+excited} classes, leading to a dataset of $N=5531$ samples, 
    \item Perform a 10-fold Speaker Independent cross-validation with one speaker as test, eight as training and one as validation set,
\end{itemize}
After reporting the results following this set up, researchers may deviate as they wish, e.g., with reporting results on the improvised part of IEMOCAP or other label subsets/merges. Numerous methods \cite{zou2022speech,wang2021fine,aftab22lightsrnet,cai21b_interspeech} have been evaluated according to this setting. For the number of folds during cross-validation, we remark that both 5-fold and 10-fold evaluations are equally reasonable choices, with the latter case using more training data per fold, hence leading to slightly higher WA and UA metrics. 


\section{Experiments to examine reproducibility}
\label{sec:Experiments}


In this section, we conduct a reproducibility study on papers that had open-sourced their implementations. Our goal is to evaluate each method (whose implementation is public) from Tables \ref{table1} and \ref{table2} according to the evaluation guidelines of \ref{eval}.  

First, we remark that, despite uploading their code, the majority of researchers omit to include checkpoints of the trained models. Notably, only two of the examined works~\cite{cai21b_interspeech,shor2022trillsson} share trained weights, which makes the reproduciblity process much easier. However, the authors of~\cite{shor2022trillsson} only share the weights of the backbone, so one has to extract the train-val-test embeddings and train a linear layer on top of them to reproduce results. Feng et al.~\cite{feng20_interspeech} have uploaded a codebase that produced errors during execution, so we could not proceed further. Aftab et al.~\cite{aftab22lightsrnet} already evaluate using the recommended guidelines, and we confirm that their reported results are reproducible, after retraining their model from scratch. Similarly, while both Zou et al.~\cite{zou2022speech} and Wang et al.~\cite{wang2021fine} do not share checkpoints, training from scratch confirms their reported results. We did not check the reproduciblity of Yang et al.~\cite{yang21c_interspeech} and Pepino et al.~\cite{pepino21_interspeech}, since their works are nearly identical with~\cite{wang2021fine}.

Next, we turn our attention to the works of \autoref{table1}. Zhu and Li \cite{zhu2022speech} propose the GLAM architecture, but their public code does not include checkpoints. Training from scratch, according to their protocol (i.e., drop the \emph{happy} class' samples) yields nearly identical results to the reported ones. However, when we attempt to retrain the model by including the data from the closely related \emph{happy} class, we obtain a significant performance drop. This is evidence that a simple assumption (here, to dismiss \emph{happy} examples) can lead to an incomplete view about a model's effectiveness. The GLAM architecture outperforms both models from \cite{xu2020improve,xu2021speech}, which follow the exact same unconventional protocol. The available code of \cite{xu2020improve,xu2021speech} was not executable or maintained. 

In \autoref{table3}, we summarize the findings from our reproducibility study, where we report the WA metric measured according to the protocol described in \ref{eval}.

\begin{table}[!ht]
    \footnotesize
    \centering
    \begin{tabular}{ccccccc}
    \toprule
       \# & Publication & Pretrained & Reproducibility & WA (\%) \\
       \cmidrule(lr){1-1} \cmidrule(lr){2-2} \cmidrule(lr){3-3}  \cmidrule(lr){4-4} \cmidrule(lr){5-5}  \cmidrule(lr){6-6}  
       1 & Cai et al.~\cite{cai21b_interspeech}      & \cmark & \cmark     & 78.15 \\
       2 & Shor and Ven.~\cite{shor2022trillsson}    & \cmark & \cmark     & 68.05 \\
       3 & Aftab et al.~\cite{aftab22lightsrnet}     & \xmark & \cmark     & 71.43 \\
       4 & Zou et al.~\cite{zou2022speech}           & \xmark & \cmark     & 69.2  \\
       5 & Wang et al.~\cite{wang2021fine}           & \xmark & \cmark     & 69.61 \\
       6 & Zhu and Li~\cite{zhu2022speech}           & \xmark & \xmark     & 63.82 \\
       7 & Feng et al.~\cite{feng20_interspeech}     & \xmark & \xmark     & -     \\
       8 & Xu et al.~\cite{xu2020improve}            & \xmark & \xmark     & -     \\
       9 & Xu et al.~\cite{xu2021speech}             & \xmark & \xmark     & -     \\
    \bottomrule
    \end{tabular}
    \caption{Reproducibility experiments. Pretrained column indicates whether the implementations share trained weights.}
    \label{table3}
\end{table}

\vspace{-0.8cm}
\section{Conclusions} \label{sec:Conclusion}

In this paper, through an empirical case review and analysis of published work of SER systems based on IEMOCAP, we investigated the choices made in the design and evaluation of SER systems through a lens of ease of comparing performance, generalizabilty and reproducibility. We encourage future works to design and evaluate their methods according to a common protocol (such as that suggested based on well executed studies). The lack of which can hinder our ability to fairly compare and reproduce results of different methods, and make collective advance as a community. In the future, we hope this case study will lead to the establishment of a well-defined set of "good practices" while designing, building and releasing data and evaluation benchmarks for SER.




\section{References}
\small
\bibliographystyle{IEEEtran}
\bibliography{main}

\begin{thebibliography}{10}
\providecommand{\url}[1]{#1}
\csname url@samestyle\endcsname
\providecommand{\newblock}{\relax}
\providecommand{\bibinfo}[2]{#2}
\providecommand{\BIBentrySTDinterwordspacing}{\spaceskip=0pt\relax}
\providecommand{\BIBentryALTinterwordstretchfactor}{4}
\providecommand{\BIBentryALTinterwordspacing}{\spaceskip=\fontdimen2\font plus
\BIBentryALTinterwordstretchfactor\fontdimen3\font minus
  \fontdimen4\font\relax}
\providecommand{\BIBforeignlanguage}[2]{{%
\expandafter\ifx\csname l@#1\endcsname\relax
\typeout{** WARNING: IEEEtran.bst: No hyphenation pattern has been}%
\typeout{** loaded for the language `#1'. Using the pattern for}%
\typeout{** the default language instead.}%
\else
\language=\csname l@#1\endcsname
\fi
#2}}
\providecommand{\BIBdecl}{\relax}
\BIBdecl

\bibitem{Schuller2013ParalinguisticsinSpeechand}
B.~Schuller, S.~Steidl, A.~Batliner, F.~Burkhardt, L.~Devillers, C.~Muller, and
  S.~S. Narayanan, ``Paralinguistics in speech and language--state-of-the-art
  and the challenge,'' \emph{Computer, Speech, and Language}, vol.~27, no.~1,
  pp. 4--39, jan 2013.

\bibitem{Mower2009Interpretingambiguousemotionalexpressions}
E.~Mower, A.~Metallinou, C.-C. Lee, A.~Kazemzadeh, C.~Busso, S.~Lee, and S.~S.
  Narayanan, ``Interpreting ambiguous emotional expressions,'' in
  \emph{Proceedings of the International Conference on Affective Computing and
  Intelligent Interaction (ACII)}, Amsterdam, The Netherlands, sep 2009.

\bibitem{iemocap}
C.~Busso \emph{et~al.}, ``{IEMOCAP:} interactive emotional dyadic motion
  capture database,'' \emph{Lang. Resour. Evaluation}, vol.~42, no.~4, pp.
  335--359, 2008.

\bibitem{aftab22lightsrnet}
A.~Aftab \emph{et~al.}, ``Light-sernet: A lightweight fully convolutional
  neural network for speech emotion recognition,'' in \emph{ICASSP 2022}.\hskip
  1em plus 0.5em minus 0.4em\relax IEEE, 2022, pp. 6912--6916.

\bibitem{zhu2022speech}
W.~Zhu and X.~Li, ``Speech emotion recognition with global-aware fusion on
  multi-scale feature representation,'' in \emph{ICASSP 2022}.\hskip 1em plus
  0.5em minus 0.4em\relax IEEE, 2022, pp. 6437--6441.

\bibitem{xu2021speech}
M.~Xu, F.~Zhang, X.~Cui, and W.~Zhang, ``Speech emotion recognition with
  multiscale area attention and data augmentation,'' in \emph{ICASSP
  2021}.\hskip 1em plus 0.5em minus 0.4em\relax IEEE, 2021, pp. 6319--6323.

\bibitem{santoso21_interspeech}
J.~Santoso \emph{et~al.}, ``{Speech Emotion Recognition Based on Attention
  Weight Correction Using Word-Level Confidence Measure},'' in \emph{Proc.
  Interspeech 2021}, 2021, pp. 1947--1951.

\bibitem{feng20_interspeech}
H.~Feng, S.~Ueno, and T.~Kawahara, ``{End-to-End Speech Emotion Recognition
  Combined with Acoustic-to-Word ASR Model},'' in \emph{Proc. Interspeech
  2020}, 2020, pp. 501--505.

\bibitem{li19n_interspeech}
Y.~Li, T.~Zhao, and T.~Kawahara, ``{Improved End-to-End Speech Emotion
  Recognition Using Self Attention Mechanism and Multitask Learning},'' in
  \emph{Proc. Interspeech 2019}, 2019, pp. 2803--2807.

\bibitem{zou2022speech}
H.~Zou, Y.~Si, C.~Chen, D.~Rajan, and E.~S. Chng, ``Speech emotion recognition
  with co-attention based multi-level acoustic information,'' in \emph{ICASSP
  2022}.\hskip 1em plus 0.5em minus 0.4em\relax IEEE, 2022, pp. 7367--7371.

\bibitem{schneider2019wav2vec}
S.~Schneider \emph{et~al.}, ``{wav2vec: Unsupervised Pre-Training for Speech
  Recognition},'' in \emph{Proc. Interspeech 2019}, 2019, pp. 3465--3469.

\bibitem{baevski2020wav2vec}
A.~Baevski \emph{et~al.}, ``wav2vec 2.0: A framework for self-supervised
  learning of speech representations,'' \emph{Advances in Neural Information
  Processing Systems}, vol.~33, pp. 12\,449--12\,460, 2020.

\bibitem{hsu2021hubert}
W.-N. Hsu \emph{et~al.}, ``Hubert: Self-supervised speech representation
  learning by masked prediction of hidden units,'' \emph{IEEE/ACM Transactions
  on Audio, Speech, and Language Processing}, vol.~29, pp. 3451--3460, 2021.

\bibitem{wang2021fine}
Y.~Wang, A.~Boumadane, and A.~Heba, ``A fine-tuned wav2vec 2.0/hubert benchmark
  for speech emotion recognition, speaker verification and spoken language
  understanding,'' \emph{arXiv preprint arXiv:2111.02735}, 2021.

\bibitem{gat2022speaker}
I.~Gat, H.~Aronowitz, W.~Zhu, E.~Morais, and R.~Hoory, ``Speaker normalization
  for self-supervised speech emotion recognition,'' in \emph{ICASSP
  2022}.\hskip 1em plus 0.5em minus 0.4em\relax IEEE, 2022, pp. 7342--7346.

\bibitem{pepino21_interspeech}
L.~Pepino, P.~Riera, and L.~Ferrer, ``{Emotion Recognition from Speech Using
  wav2vec 2.0 Embeddings},'' in \emph{Proc. Interspeech 2021}, 2021, pp.
  3400--3404.

\bibitem{yang21c_interspeech}
S.~wen Yang \emph{et~al.}, ``{SUPERB: Speech Processing Universal PERformance
  Benchmark},'' in \emph{Proc. Interspeech 2021}, 2021, pp. 1194--1198.

\bibitem{cai21b_interspeech}
X.~Cai, J.~Yuan, R.~Zheng, L.~Huang, and K.~Church, ``{Speech Emotion
  Recognition with Multi-Task Learning},'' in \emph{Proc. Interspeech 2021},
  2021, pp. 4508--4512.

\bibitem{shor2022universal}
J.~Shor, A.~Jansen, W.~Han, D.~Park, and Y.~Zhang, ``Universal paralinguistic
  speech representations using self-supervised conformers,'' in \emph{ICASSP
  2022}.\hskip 1em plus 0.5em minus 0.4em\relax IEEE, 2022, pp. 3169--3173.

\bibitem{gulati2020conformer}
A.~Gulati \emph{et~al.}, ``Conformer: Convolution-augmented transformer for
  speech recognition,'' \emph{arXiv preprint arXiv:2005.08100}, 2020.

\bibitem{shor20_interspeech}
J.~Shor \emph{et~al.}, ``{Towards Learning a Universal Non-Semantic
  Representation of Speech},'' in \emph{Proc. Interspeech 2020}, 2020, pp.
  140--144.

\bibitem{shor2022trillsson}
J.~Shor and S.~Venugopalan, ``Trillsson: Distilled universal paralinguistic
  speech representations,'' \emph{arXiv preprint arXiv:2203.00236}, 2022.

\bibitem{musgrave2020reality}
K.~Musgrave, S.~J. Belongie, and S.~Lim, ``A metric learning reality check,''
  in \emph{ECCV 2020}.

\bibitem{lotfian2017building}
R.~Lotfian and C.~Busso, ``Building naturalistic emotionally balanced speech
  corpus by retrieving emotional speech from existing podcast recordings,''
  \emph{IEEE Transactions on Affective Computing}, vol.~10, no.~4, pp.
  471--483, 2017.

\bibitem{Chatziagapi2019}
A.~Chatziagapi \emph{et~al.}, ``{Data Augmentation Using GANs for Speech
  Emotion Recognition},'' in \emph{Proc. Interspeech 2019}, 2019, pp. 171--175.

\bibitem{pappagari2021copypaste}
R.~Pappagari, J.~Villalba, P.~Żelasko, L.~Moro-Velazquez, and N.~Dehak,
  ``Copypaste: An augmentation method for speech emotion recognition,'' in
  \emph{ICASSP 2021}.\hskip 1em plus 0.5em minus 0.4em\relax IEEE, 2021, pp.
  6324--6328.

\bibitem{xu2020improve}
M.~Xu, F.~Zhang, and S.~U. Khan, ``Improve accuracy of speech emotion
  recognition with attention head fusion,'' in \emph{2020 10th Annual Computing
  and Communication Workshop and Conference (CCWC)}, 2020, pp. 1058--1064.

\bibitem{peng2021efficient}
Z.~Peng, Y.~Lu, S.~Pan, and Y.~Liu, ``Efficient speech emotion recognition
  using multi-scale cnn and attention,'' in \emph{ICASSP 2021}.\hskip 1em plus
  0.5em minus 0.4em\relax IEEE, 2021, pp. 3020--3024.

\bibitem{liu21n_interspeech}
J.~Liu and H.~Wang, ``{A Speech Emotion Recognition Framework for Better
  Discrimination of Confusions},'' in \emph{Proc. Interspeech 2021}, 2021, pp.
  4483--4487.

\bibitem{moine21_interspeech}
C.~L. Moine, N.~Obin, and A.~Roebel, ``{Speaker Attentive Speech Emotion
  Recognition},'' in \emph{Proc. Interspeech 2021}, 2021, pp. 2866--2870.

\bibitem{chernykh2017emotion}
V.~Chernykh and P.~Prikhodko, ``Emotion recognition from speech with recurrent
  neural networks,'' \emph{arXiv preprint arXiv:1701.08071}, 2017.

\bibitem{tarantino19_interspeech}
L.~Tarantino, P.~N. Garner, and A.~Lazaridis, ``{Self-Attention for Speech
  Emotion Recognition},'' in \emph{Proc. Interspeech 2019}, 2019, pp.
  2578--2582.

\end{thebibliography}
\end{document}